\documentclass[aps,preprint]{revtex4}%
\usepackage{amsfonts}
\usepackage{amsmath}
\usepackage{amssymb}
\usepackage{graphicx}%
\begin{document}
\preprint{ }
\title[Lasers and proxies]{Laser fields and proxy fields}
\author{H. R. Reiss}
\affiliation{Max Born Institute, 12489 Berlin, Germany}
\affiliation{American University, Washington, DC 20016-8058, USA}
\email{reiss@american.edu}


\begin{abstract}
The convention in Atomic, Molecular, and Optical (AMO) physics of employing
the dipole approximation to describe laser-induced processes replaces four
source-free Maxwell equations governing laser fields with a single Maxwell
equation for a \textquotedblleft proxy\textquotedblright\ field that requires
a virtual source current for its existence. Laser fields are transverse, but
proxy fields are longitudinal; there can be no gauge equivalence. The proxy
field is sometimes serviceable, but its limitations are severe. One example is
the \textquotedblleft above-threshold ionization\textquotedblright\ (ATI)
phenomenon; surprising by proxy-field understanding, but natural and predicted
in advance of observation with a laser-field method. An often-overlooked
limitation is that numerical solution of the time-dependent Schr\"{o}dinger
equation (TDSE) is exact for proxy fields, but not for laser fields.
Acceptance of proxy-field concepts has been costly in terms of inefficiently
deployed research resources. Calculations with a nearly-40-year old
transverse-field method remain unmatched with proxy fields. The
transverse-field method is applicable in the \textquotedblleft
tunneling\textquotedblright\ domain, the \textquotedblleft
multiphoton\textquotedblright\ domain, and, as shown here, in the
low-frequency \textquotedblleft magnetic\textquotedblright\ domain. Attempts
to introduce low-frequency magnetic field corrections into TDSE cannot be
expected to produce meaningful results. They would be based on inappropriate
Maxwell equations, a non-existent virtual source, and would approach constant
electric field properties as the field frequency declines. Laser fields
propagate at the speed of light for all frequencies; they cannot approach a
constant-field limit. Extremely strong laser fields are unambiguously
relativistic; a nonrelativistic limit that connects continuously to the
relativistic domain is simpler conceptually and mathematically than is a
theory constructed with a proxy field that is certain to fail as intensities increase.

\end{abstract}
\date[21 February 2017]{}
\maketitle

Strong-field laser interactions with matter in the nonrelativistic domain are
conventionally treated within the dipole approximation. The investigation
carried out here begins with a comparison of the Maxwell equations governing
actual laser fields with those relating to the dipole approximation. These
equations are so different, mathematically and in their physical significance,
that the terminology is adopted of referring to the dipole-approximate version
as a \textit{proxy field} employed to represent an actual laser field. The
task is then to examine how well the proxy field mimics the actual laser
field. The ability of the proxy field to match successfully some laboratory
phenomena means that the mimicry can be satisfactory in those cases, but it is
also found that the mimicry is not merely bounded in extent but it can be
seriously in error in important ways. A basic difficulty with the proxy field
is that it cannot exist without a postulated virtual source current. This
contrasts with laser fields that, after their initial creation, propagate in
vacuum indefinitely without sources to sustain them. The virtual source
current is necessary for mimicry, but it can also introduce unwanted and
unexpected consequences, one of which is the unphysical insertion of external
energy and momentum if the virtual source current is employed incautiously.
Some incorrect results from faulty mimicry have been accepted as valid in the
strong-field community, such as a zero-frequency limit of a constant electric
field, and a reliance on the numerical solution of the Schr\"{o}dinger
equation for a proxy field as being exact. These are basic qualitative
problems with proxy fields. Quantitative shortcomings of proxy fields are
demonstrated by means of 30-year-old transverse-field comparisons with
experiments that have yet to be matched by proxy calculations either in
precision or in universality. The implication is that the use of the dipole
approximation has severely hampered progress in the understanding of
strong-field laser interactions with matter, and has thereby caused an
inefficient expenditure of valuable research resources. Introduction of
low-frequency corrections into TDSE, projected by some investigators, would
only continue the misdirection of resources, since such an extension applies
only to proxy fields, whereas magnetic-field effects are already present in a
transverse-field method, as exhibited below.

The dipole approximation applied to laser fields amounts to replacing a
transverse field by a longitudinal field. This approximation has been so
successful in perturbative Atomic, Molecular, and Optical (AMO) physics that
the education of AMO practitioners is based almost entirely on
longitudinal-field effects, with little awareness of the basic differences
with transverse fields, even though those differences are essential at high
intensities and/or low frequencies \cite{hr101}.

The dipole-approximation approach to strong fields began with the Keldysh
theory of 1964 \cite{keldysh}, and was elaborated by other long-standing works
\cite{nikrit,ppt,adk}. Some criteria based upon dipole-only\ concepts have
been applied even in the relativistic environment. It is shown herein that
comparison with experiments of a transverse-field theory published long ago
\cite{hjhr,hr80} has yet to be matched in accuracy or universality by
proxy-field methods, almost 40 years later. The present article is an attempt
to provide an effective and efficient focus for future work.

The change in approach to strong-field phenomena proposed here amounts to a
change of paradigm (in the sense defined by Thomas S. Kuhn \cite{kuhn}) for
the study of the effects of very strong laser fields on matter
\cite{hrparadigm}. When the laser field is represented properly as a
transverse field, physical interpretations differ from those arrived at in the
dipole approximation, and calculated results are also different. It is shown
below that both qualitatively and quantitatively, the transverse-field
approach gives correct insights and accurate numerical modelings that are
unmatched by the proxy field that follows from the dipole approximation. One
fundamental example is ATI, that created a crisis for dipole-approximation
methods that took years to resolve, whereas it was predicted in advance of
observation with transverse-field methods. A major problem that looms for
strong-field physics is that many investigators are proposing to insert
magnetic-field corrections into existing numerical methods (known as TDSE for
\textit{time-dependent Schr\"{o}dinger equation}) that are based on the dipole
approximation; but such a correction to the proxy field does not provide the
necessary bridge to transverse-field effects because it depends on the Maxwell
equation for the proxy field, and it retains the virtual source current
required for the proxy field. It would also maintain the concept of
\textit{adiabaticity}, the notion that low frequencies imply a progression to
static-field behavior, which is unphysical because laser fields propagate with
the speed of light for all frequencies and cannot have a static limit.

The vacuum Maxwell equations in Gaussian units for the electric field
$\mathbf{E}$ and the magnetic field $\mathbf{B}$ are%
\begin{align}
\mathbf{\nabla\cdot E}  &  =4\pi\rho,\label{a}\\
\mathbf{\nabla\cdot B}  &  =0,\label{b}\\
\mathbf{\nabla\times B}-\frac{1}{c}\partial_{t}\mathbf{E}  &  =\frac{4\pi}%
{c}\mathbf{J,}\label{c}\\
\mathbf{\nabla\times E}+\frac{1}{c}\partial_{t}\mathbf{B}  &  =0, \label{d}%
\end{align}
where $\rho$ and $\mathbf{J}$ are the source charge and current densities. As
applicable to any transverse field such as that of a laser, these equations
apply with no source terms, so that laser fields are governed by the equations%
\begin{align}
\mathbf{\nabla\cdot E}  &  =0,\label{e}\\
\mathbf{\nabla\cdot B}  &  =0,\label{f}\\
\mathbf{\nabla\times B}-\frac{1}{c}\partial_{t}\mathbf{E}  &  =0,\label{g}\\
\mathbf{\nabla\times E}+\frac{1}{c}\partial_{t}\mathbf{B}  &  =0. \label{h}%
\end{align}
The symmetry between electric and magnetic fields that is evident in these
equations is a hallmark of transverse fields. An important reminder is that,
in any laser experiment, the only fields that can actually reach the target
are propagating fields. (The descriptions \textit{transverse},
\textit{propagating}, \textit{plane-wave} and \textit{laser }are used
interchangeably here.) Any extraneous fields introduced by interactions with
optical elements cannot persist beyond a few wavelengths from those
disturbances. Thus, the beam reaching the target can consist only of fields
that obey the plane-wave equations (\ref{e}) to (\ref{h}).

The dipole approximation (DA), as conventionally employed in the AMO
community, is defined by the substitutions%
\begin{equation}
\mathbf{E}\left(  t,\mathbf{r}\right)  \rightarrow\mathbf{E}^{DA}\left(
t\right)  ,\quad\mathbf{B}\left(  t,\mathbf{r}\right)  \rightarrow
\mathbf{B}^{DA}=0. \label{i}%
\end{equation}
The applicable Maxwell equations follow from zero values for all expressions
in (\ref{a}) - (\ref{d}) containing the $\mathbf{\nabla}$ operator, and the
complete absence of $\mathbf{B}$. With these limitations, the only surviving
equation is (\ref{c}), which can be written as%
\begin{equation}
\partial_{t}\mathbf{E}^{DA}\left(  t\right)  =-4\pi\mathbf{J}^{DA}\left(
t\right)  , \label{j}%
\end{equation}
where $\mathbf{J}^{DA}\left(  t\right)  $ is a virtual source current that
must be posited in order to satisfy the Maxwell equation applicable to the
dipole approximation. The fidelity of dipole-approximation mimicry to actual
laser fields is determined by how well the effects of the proxy field governed
by Eq. (\ref{j}) replicate those of laser fields in Eqs. (\ref{e}) - (\ref{h}).

An example of effective mimicry (with important limitations) is the matter of
\textit{rescattering}, a process in which an electron ionized from an atom
follows a path that returns it to the ion, upon which it can recombine with
the accompanying emission of higher harmonics of the laser field. The
classical path of a free electron in the laser field \cite{ss} matches, in the
nonrelativistic case, exactly that of the electron driven by the virtual
source $\mathbf{J}^{DA}$ \cite{kuchiev,kulander,corkum}, as long as the
comparison is limited to not much more than a wavelength in space or a wave
period in time. Beyond those narrow limits, the virtual source current injects
unphysical energy into the problem, resulting in a photoelectron trajectory
that fails to mimic the motion of a free electron in the laser field. The
rescattering case is especially instructive because it illustrates that
effective mimicry evokes models that are completely different from those
appropriate to laser fields. The proxy field forces an oscillatory electron
motion, called a \textit{quiver motion}, with a kinetic energy termed a
\textit{ponderomotive energy}. In the free electron case, the same magnitude
of ponderomotive energy is involved, but it is not kinetic; it arises from a
\textquotedblleft mass shift\textquotedblright\ of a charged particle in a
plane-wave field \cite{hrup}.

An example is provided by the path of a photoelectron after ionization by a
circularly polarized field. According to a tunneling theory \cite{corkum}, the
photoelectron follows a trajectory described by the coordinates%
\begin{align}
x  &  =x_{0}\left(  1-\cos\omega t\right)  ,\label{j1}\\
y  &  =x_{0}\left(  \omega t-\sin\omega t\right)  , \label{j2}%
\end{align}
where $x_{0}=E_{0}/\omega^{2}$, $E_{0}$ is the amplitude of the oscillatory
electric field of frequency $\omega$, and the $x$ and $y$ coordinate axes
define a plane perpendicular to the propagation direction of the laser field.
The anomaly is most easily seen from the history of the angular momentum of
the photoelectron as it follows the trajectory given by Eqs. (\ref{j1}) and
(\ref{j2}). After the first few cycles, the angular momentum $l$ has the
simple description \cite{hrjpb}%
\begin{equation}
l=\omega^{2}x_{0}^{2}t\sin\omega t, \label{j3}%
\end{equation}
which oscillates in sign and increases in amplitude linearly with time. This
is completely unphysical for at least two reasons. First is the simple fact
that a circularly polarized field can transfer only one sign of angular
momentum to the photoelectron; oscillation in sign is qualitatively
impossible. There is also the fact that the only angular momentum that can be
imparted to the photoelectron corresponds to the number of photons absorbed in
the ionization process, with each photon contributing one quantum unit of
angular momentum. An angular momentum of the photoelectron increasing
indefinitely with time is a quantitative impossibility. Both of these
anomalies arise from the virtual current $\mathbf{J}^{DA}$, which continues to
transfer energy and momentum to the photoelectron long after the brief time
allowed by any reasonable mimicry requirement. A further problem is that the
relations (\ref{j1}) and (\ref{j2}) depend on initial conditions applied at
the \textquotedblleft tunnel exit\textquotedblright. A tunneling event is
possible only from the interaction of two scalar fields, which can happen only
if the laser field is treated within the dipole approximation.

The combination of an actual laser field, which is a vector quantity, with a
scalar Coulomb potential, cannot be represented by a tunneling model.

A very important example of failed mimicry is the requirement that the
zero-frequency limit of the laser field should match known behavior of systems
in a constant electric field. This is a generally accepted condition,
exemplified by Refs. \cite{joachain,jbauer}. It is a natural criterion when
the dipole-approximation $\mathbf{r\cdot E}$ interaction Hamiltonian is
applicable, but it is critically incorrect for laser fields. Laser fields
propagate with the speed of light in vacuum for any frequency, no matter how
low it is. A propagating field cannot have a static field as a limit, and
transverse fields differ most radically from longitudinal fields at low
frequencies \cite{hr101}.

One fact immediately evident from comparing the Maxwell equations (\ref{e}) -
(\ref{h}) for a laser field with (\ref{j}) for the proxy field is that there
can be no gauge equivalence connecting them. Gauge transformations preserve
the fields, and it is obvious that the laser field and the proxy field are not
the same. All tunneling theories \cite{keldysh,nikrit,ppt,adk,wb} are based on
the dipole-approximation Maxwell equation (\ref{j}). One basis for the
conclusion that proxy theories are superior to a transverse-field theory
\cite{dbauer} comes from comparing the analytical approximations with TDSE
numerical solutions. The TDSE result is considered to be exact, although it is
actually based on the proxy-field theory of (\ref{j}). The reasoning is
circular; all that is shown is that analytical approximations arising from the
DA agree with numerical solution of the Schr\"{o}dinger equation arising from
the DA. The fact that these results disagree with the laser-field results of
Ref. \cite{hr80} is a judgment to the detriment of the proxy field, rather
than in favor of the proxy theory as concluded in Ref. \cite{dbauer}.

Clarifying remarks about the foundations of the 1980 theory \cite{hr80} are
appropriate here. A completely relativistic theory of strong-field ionization
was developed in 1990, based on the Volkov solution, which is an exact
solution of the Dirac equation for the behavior of a charged particle in a
laser field. When the nonrelativistic limit of this theory is taken, the
result so obtained is the 1980 theory \cite{hr80,hr90,hrrel,dpcdiss}. The 1980
theory thus includes the effects of the magnetic component of the laser field
since the long-wavelength limit is taken only \textit{after} integration over
the spatial part of the full $\left(  \omega t-\mathbf{k\cdot r}\right)  $
phase of the laser field has been accomplished. This contrasts with the
\textit{ab initio} use of the dipole approximation in the proxy theory. See
the Supplemental Material \cite{suppl} for more information.

A problem of perception arises from the regrettable acceptance of the acronym
KFR. The Keldysh (K) theory \cite{keldysh} is a dipole-approximation theory
expressed in the length gauge and applicable in principle only to low
frequencies since it depends on the tunneling concept. The Faisal (F) theory
\cite{faisal} is also a dipole-approximation theory, but expressed in the
velocity gauge and based on an approximation valid only for high frequencies.
The theory described as \textquotedblleft R\textquotedblright\ \cite{hr80} is
the only component of KFR that is formulated for transverse fields, it is
expressed in the radiation gauge, and it does not have inherent frequency
limitations. \textquotedblleft KFR\textquotedblright\ is confusing and
unfortunate. The terminology Strong-Field Approximation (SFA) was introduced
in 1990 \cite{hr90} to distinguish the laser-field theory from proxy-field
theories, but this attempt at clarity was nullified by the adoption in 1994
\cite{lewenstein} of \textquotedblleft SFA\textquotedblright\ to label all
analytical approximations.

A fundamental problem arising from the contrasting Maxwell equations for laser
fields and proxy fields is the matter of nondipole corrections. It has been
pointed out \cite{hr101} that the well-known failure of the dipole
approximation at very high frequencies is overshadowed by the little-known but
very important failure of the dipole approximation at low frequencies, brought
on by the increasing importance of the magnetic component of a laser field as
the frequency declines. It has been suggested by specialists in TDSE
calculations that an apparently straightforward solution of the problem is to
incorporate nondipole corrections into the proxy-field Schr\"{o}dinger
equation. However, no alteration of Eq. (\ref{j}) can replicate Eqs. (\ref{e})
- (\ref{h}). The result of nondipole corrections in a theory based on the
proxy Maxwell equation (\ref{j}) would serve only to provide information about
longitudinal waves \cite{wesley}, a little-known phenomenon unrelated to laser fields.

When a laser field, characterized by its unique ability to propagate
indefinitely without the benefit of charge or current sources, is modeled by a
proxy field completely devoid of a magnetic component and dependent for its
existence on a virtual current, it is inevitable that qualitative
understanding of laser-induced phenomena can be importantly different in the
two cases. One instance of this has already been noted in the matter of the
ponderomotive energy, which is a kinetic energy for the proxy field and more
in the nature of a potential energy for the laser field. Another example of
basic importance is the Above-Threshold Ionization (ATI) phenomenon
\cite{ati}, wherein the long-familiar dominance of the lowest-order process in
perturbation theory is replaced by a concept in which many orders can
contribute, and where the lowest order might not even be the most important.
The proxy theory, with its dependence on a single Maxwell equation, appears to
be governed by a single parameter, the Keldysh parameter $\gamma_{K}$, defined
as%
\begin{equation}
\gamma_{K}=\sqrt{E_{B}/2U_{p}}, \label{k}%
\end{equation}
where $E_{B}$ is the energy by which an electron is bound in an atom, and
$U_{p}$ is the already-discussed ponderomotive energy of the free electron in
the field. This leads to a physical model divided into two domains, where the
low-frequency \textit{tunneling} domain ($\gamma_{K}<1$) shows no clearly
distinguished peaks in photoelectron spectra, and the higher-frequency
\textit{multiphoton} domain ($\gamma_{K}>1$) reveals individual peaks in a
spectrum that are identifiable with specific numbers of photons that
participate in the process. Tunneling is an explicitly proxy-field process.
There is no such concept as tunneling in a laser-field theory. Two separate
intensity parameters are required for the laser-field theory, conveniently
expressed in terms of the ratio of the basic ponderomotive energy to the
energy of a photon and to the binding energy in the atom, and expressed as
\cite{hr80}%
\begin{equation}
z=U_{p}/\hbar\omega,\quad z_{1}=2U_{p}/E_{B}. \label{l}%
\end{equation}
The $z_{1}$ parameter is related to the Keldysh parameter ($z_{1}=1/\gamma
_{K}^{2}$), but nothing equivalent to $z$ exists for the proxy field.

These manifest differences in the basic descriptions of laser fields and proxy
fields are related to the physical interpretation of the ATI phenomenon. The
first observation of ATI \cite{ati} was regarded as a shocking and unexpected
development within the DA-dominated AMO community, but it was obvious and
predicted in full detail in advance of the laboratory observation within a
laser-field treatment. In fact, important aspects of the ATI phenomenon that
were not observed until 1986 \cite{bucks86} were already predicted by the
theory published in 1980 \cite{hr80}. The 1980 theory was created and
discussed prior to 1979. (See also Ref. \cite{hjhr}, where some ATI features
are described.) Analogs of ATI were predicted and commented upon as long ago
as 1970 \cite{hr70,hr71}. That is, ATI was predicted in advance of its initial
observation. This was possible because a theory of transverse fields was employed.%

\begin{figure}
[ptb]
\begin{center}
\includegraphics[
height=2.8784in,
width=6.7837in
]%
{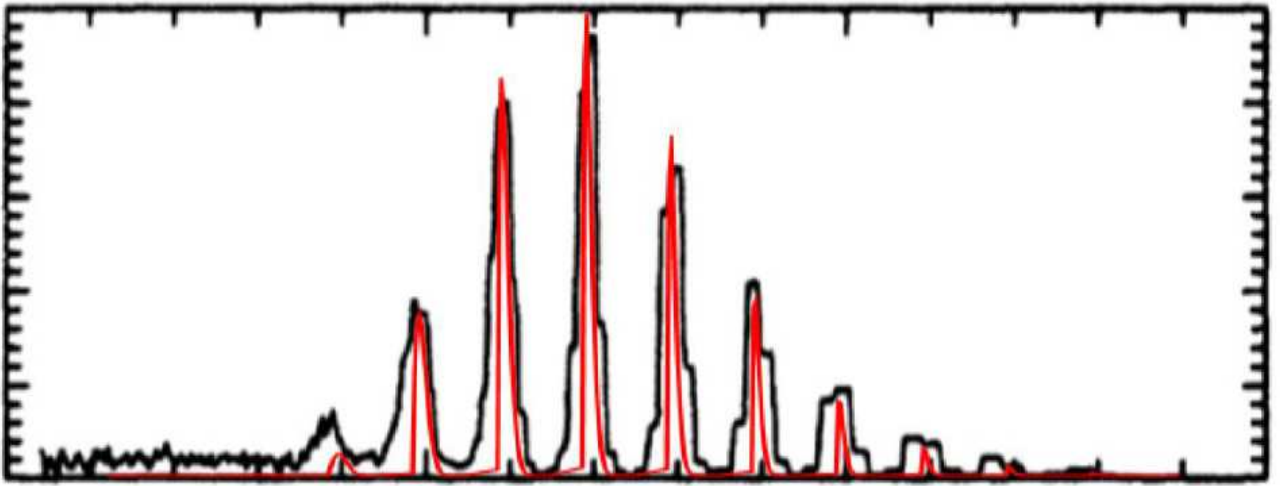}%
\caption{(color online) This figure shows the ability of the transverse-field
theory of Ref. \cite{hr80} to replicate the experimental results presented in
Ref. \cite{bucks86}. In DA terminology, this is in the multiphoton domain. The
black curve (with wide peaks) is the measured photoelectron spectrum and the
red curve (with narrow peaks) is the theoretical fit. Laser parameters:
$1064nm$, peak intensity $2\times10^{13}W/cm^{2}$, pulse duration $100ps$
($z=1.82,$ $z_{1}=0.35,$ $\gamma_{k}=1.69$) on a xenon target. The calculation
includes focal averaging in a Gaussian beam with Gaussian time distribution,
and with partial ponderomotive energy ($U_{p}$) recovery in the very long
pulse. The only fitting parameter employed was the relative fraction of
recovered $U_{p}$ (about $80\%$) selected to fix the absolute energy locations
of the peaks. The theory used was in existence prior to the first observation
of ATI \cite{ati}. It is believed that no other theory or TDSE calculation can
match this correspondence between theory and experiment in this parameter
domain.}%
\label{fig1}%
\end{center}
\end{figure}

\begin{figure}
[ptb]
\begin{center}
\includegraphics[
height=4.3653in,
width=3.946in
]%
{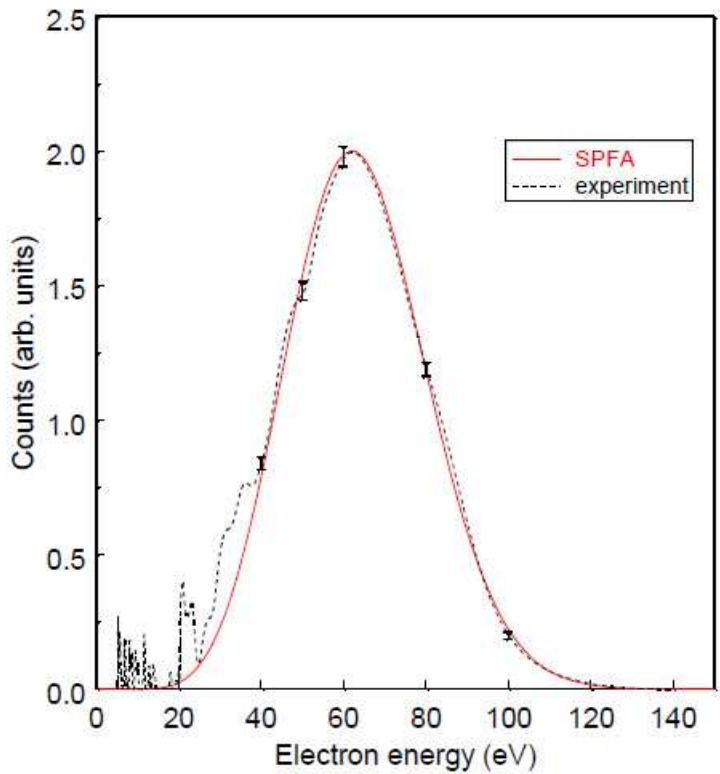}%
\caption{(color online) This figure, based on Ref. \cite{hr96}, shows the
extreme accuracy possible with the transverse-field theory of 1980 \cite{hr80}
applied to the description of an experiment \cite{moh} at the relatively high
intensity of $1.27\times10^{15}W/cm^{2}$ at $815nm$ wavelength and a pulse
length of $180fs$ in the ionization of helium ($z=52,$ $z_{1}=6.4,$
$\gamma_{K}=0.40$). By DA terminology, this is in the tunneling domain. The
fit is within the very small experimental error bars. Irregularities in the
low energy part of the spectrum are experimental artifacts \cite{mohpriv}. The
label \textquotedblleft SPFA\textquotedblright\ stands for \textquotedblleft
Strong Propagating-Field Approximation\textquotedblright\ to distinguish it
from the ambiguous \textquotedblleft SFA\textquotedblright.}%
\label{fig2}%
\end{center}
\end{figure}

Figure \ref{fig1} is shown here for several reasons. The experimental data
plotted in the figure are from the 1986 experiments by Bucksbaum, \textit{et
al. }\cite{bucks86}. A 1987 paper \cite{hr87} based on the 1980 theory
correctly assigned the relative probabilities of the ATI peaks. This result
was then extended by Bucksbaum, shown at a 1988 conference \cite{buckskos}, to
include averaging over the distribution of intensities in the laser focus.
Figure \ref{fig1} shows the result of a recent detailed recalculation, based
entirely on the 1980 paper, including averaging over the spatial and temporal
variations of laser intensity in the focal region, and also including partial
return of ponderomotive energy to the photoelectrons emerging from the very
long $100ps$ pulse duration. The best estimate for peak laser intensity
provided by one of the authors \cite{mcilrath} of Ref. \cite{bucks86} is
employed, and the only adjustable parameter is the relative fraction of
$U_{p}$ returned to the photoelectron upon leaving the pulse. This parameter
establishes the absolute energy location of each of the ATI peaks. These
results from a 1986 experiment, described by a 1980 transverse-field theory,
have yet to be matched by any proxy-field calculation even 30 years later.
Another example of unmatched agreement between experimental measurements and
theoretical predictions is with results from a 1993 experiment \cite{moh},
replicated by a 1996 transverse-field calculation \cite{hr96}, is shown in
Fig. \ref{fig2}.%

A recent example of the advantages in physical interpretation provided by a
transverse-field explanation comes from very precise experiments on the
displacement in the propagation direction of photoelectrons generated by a
circularly polarized laser \cite{smeenk}. Performed with nonrelativistic laser
fields of wavelengths of $800nm$ and $1400nm$, the very small effect was
nevertheless detected and identified by the authors as due to radiation
pressure. From the point of view of Eqs. (\ref{e}) - (\ref{h}), it is
elementary to understand the magnitude of the effect and the fact that, in
strong fields, the effect is independent of the identity of the parent atom
\cite{hrsmeenk}. With circular polarization, the most probable energy required
from the laser beam for ionization is $E_{B}+2U_{p}$, where $E_{B}$ is the
binding energy of the electron in the atom and $U_{p}$ is the ponderomotive
energy of the free electron in the laser field. Of this, $E_{B}+U_{p}$ is
required to reach ionization threshold, and the associated photon momentum is
transferred to the entire atom, with negligible consequences. The remaining
energy $U_{p}$ is necessary to account for the kinetic energy of the
photoelectron in a classical orbit around the ion. The associated photon
momentum of $U_{p}/c$ is transferred to the photoelectron, which explains
exactly the behavior found in the experiments. The experimenters sought
explanation in a DA theory \cite{smeenk,cbc1,cbc2}. They experienced daunting
problems that have a simple origin: an electric field alone can only provide
forces in the polarization direction of the electric field, whereas radiation
pressure is exerted in the direction of propagation. The combined action of
electric and magnetic fields is necessary to explain a force in the
propagation direction. The proxy-field theory, as shown in Eq. (\ref{i}), has
no magnetic component at all.

An important example showing the low-frequency onset of magnetic field effects
comes from 1988 experiments \cite{laval} with a $CO_{2}$ laser operating at
about $10\mu m$ at intensities up to $10^{14}W/cm^{2}$. This is a wavelength
that, from the proxy-field point of view, is clearly in the tunneling domain.
From a laser-field point of view, the experiments fall in the region where
magnetic field effects ($v/c$ effects) become important, but purely
relativistic effects ($\left(  v/c\right)  ^{2}$ effects) can be neglected
\cite{hr101}. The 1988 experiments with xenon ionized by a linearly polarized
laser field showed a very sharp peak in the spectrum at low energy, followed
by a second strong peak at about $150eV$. This cannot be explained by a
tunneling theory, but it matches precisely the predictions of a
transverse-field theory \cite{hr102}. Both peaks are explicable by properties
of the momentum-space wave function of xenon.%

\begin{figure}
[ptb]
\begin{center}
\includegraphics[
trim=0.000000in 1.202686in 0.000000in 0.963299in,
height=4.0498in,
width=7.7102in
]%
{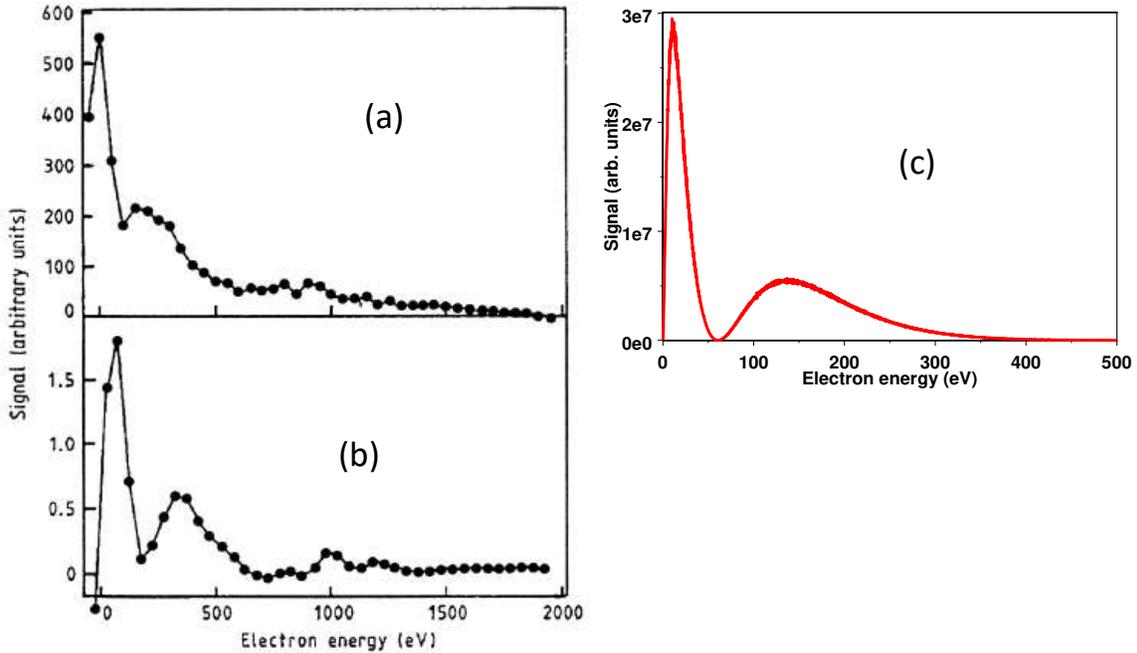}%
\caption{(color online) Part (a) is a reproduction from Ref. \cite{laval} of
experimental spectra of photoelectrons emitted from xenon with a linearly
polarized laser operating at $10.6\mu m$ at an intensity of $10^{14}W/cm^{2}.$
The double peak was so unexpected that the authors tried to see if it could be
eliminated by a \textquotedblleft smoothing\textquotedblright\ algorithm,
leading to part (b) with multiple peaks still present. The Keldysh parameter
is $\gamma_{K}=0.077$, which is deeply within the tunneling regime in a proxy
field context. It is within the \textquotedblleft magnetic-field
effects\textquotedblright\ region in a transverse-field theory \cite{hr101}. A
tunneling theory cannot explain the secondary peak at about $150eV$. Part (c)
replicates Fig. 4(b) in Ref. \cite{hr102}, calculated by a transverse-field
theory \cite{hr80} from the stated parameters in Ref. \cite{laval} with no
fitting done. It matches very well the location of the two prominent peaks in
part (a). (Note the difference in the x-axis scales, that go only to $500eV$
in part (c).)}%
\label{fig3}%
\end{center}
\end{figure}

In summary, laser effects arise from transverse fields while DA proxy fields
are longitudinal. Therefore, proxy fields obey completely different Maxwell
equations, lose the ability to explain source-free propagation, and therefore
provide qualitative explanations that may not be in accord with laboratory
reality. The proxy fields require a virtual source current that can inject
unphysical energy into a problem if necessary precautions are not observed.
Transverse-field strong-field theories \cite{hjhr,hr80,hrrel} have been in
existence for a long time, they provide clear physical explanations for
strong-laser effects, but further development of these long-standing theories
has been neglected in favor of the seemingly simpler dipole-approximation
theories that have almost exclusively engaged the attention of strong-field
laser research. The defects of the proxy-field approach become severe at very
low frequencies, and relief from these difficulties are not to be found in
corrections to DA results because the Maxwell equations and the
Schr\"{o}dinger equation involved are inappropriate for the task. Extensions
of currently available theories that go beyond the tunneling model can be
found from Eqs. (\ref{e}) - (\ref{h}), based on simplified versions of the
relativistic theories of Refs. \cite{hrrel,dpcdiss}. Breadth of applicability
of a transverse-field theory is shown by Fig. \ref{fig1} for the DA-labeled
multiphoton domain, Fig. \ref{fig2} for the tunneling domain of the DA, and
Fig. \ref{fig3} for the magnetic domain.

Extremely strong laser fields are relativistic; a nonrelativistic limit that
connects continuously to the relativistic domain is simpler conceptually and
mathematically and is more immune to unexpected pitfalls than is a theory
constructed with a proxy field that is certain to fail as intensities increase.

The overall conclusion: the \textquotedblleft bottom line\textquotedblright,
is that human and financial resources expended on the study of strong-laser
effects by a theory founded on a proxy field, to the neglect of approaches
based on actual laser fields, are resources expended on insecure grounds.
After nearly four decades of effort, dipole-approximation theories still
cannot match the accuracy and universality of those based on actual laser
fields, and present the constant threat of inappropriate physical insights and
inefficient expenditure of limited research resources.

\end{document}


\preprint{ }
\title[Supplementary material]{Supplemental material for\\Laser fields and proxy fields}
\author{H. R. Reiss}
\affiliation{Max Born Institute, 12489 Berlin, Germany}
\affiliation{American University, Washington, DC 20016-8058, USA}
\maketitle

The procedure shown here is for a nonrelativistic environment, but all of the
results have been shown \cite{hr90,hrrel} to be the nonrelativistic spinless
limit of a relativistic Dirac theory. The nonrelativistic limit is
demonstrated \textit{after} integration over the spatial coordinate dependence
of the complete phase $\omega t-\mathbf{k\cdot r}$ has been accomplished. The
significance of this is that magnetic field effects, of order $v/c$ as
compared to electric field effects, are retained even in the nonrelativistic
limit. Relativistic corrections are of order $\left(  v/c\right)  ^{2}$.

A full development of the theory is to be found in Ref. \cite{hr80}, but that
theory is stated in \textquotedblleft natural units\textquotedblright%
\ ($\hbar=1$, $c=1$) as employed in relativistic strong-field theory. What
follows is in atomic units, more suited to the nonrelativistic environment.

The theory employed is an S-matrix theory that is based on the way in which
laboratory measurements are related to a theoretical description. Although the
formalism is exact, it is usually necessary to introduce some physically
motivated approximation for at least one element in the formal statement. Two
equivalent forms are possible: a direct-time form, and a time-reversed form.
The direct-time form of the transition amplitude is%
\begin{equation}
\left(  S-1\right)  _{fi}=-i\int_{-\infty}^{\infty}dt\left(  \Phi_{f}%
,H_{I}\Psi_{i}\right)  , \label{a}%
\end{equation}
and the time-reversed form is%
\begin{equation}
\left(  S-1\right)  _{fi}=-i\int_{-\infty}^{\infty}dt\left(  \Psi_{f}%
,H_{I}\Phi_{i}\right)  , \label{b}%
\end{equation}
where $\Phi$ is a solution of the Schr\"{o}dinger equation lacking the term
representing the interaction that causes the transition, and $\Psi$ is a
solution of the Schr\"{o}dinger equation in which all influences are included:%
\begin{align}
i\partial_{t}\Phi &  =H_{0}\Phi,\label{c}\\
i\partial_{t}\Psi &  =\left(  H_{0}+H_{I}\right)  \Psi. \label{d}%
\end{align}
The indices $i$ and $f$ in Eqs. (\ref{a}) and (\ref{b}) stand for
\textquotedblleft initial\textquotedblright\ and \textquotedblleft
final\textquotedblright. Each of these forms includes one fully interacting
state $\Psi$ and one non-interacting or \textquotedblleft
reference\textquotedblright\ state $\Phi$. For strong-field atomic ionization,
the time-reversed form (\ref{b}) is preferred, since the initial state can be
represented precisely by an exact solution of the Schr\"{o}dinger equation if
it is known, or otherwise by an analytical Hartree-Fock solution that can be
very accurate. When ionization is caused by an intense laser field, the final
state $\Psi_{f}$ can be approximated by a Volkov solution, which is an exact
solution of the Dirac equation for a free electron in a plane-wave field of
arbitrarily high intensity, consisting of any distribution of co-propagating
frequency components. The initial bound state is just a stationary state, so
that%
\begin{equation}
\Phi_{i}\left(  t,\mathbf{r}\right)  =\phi_{i}\left(  \mathbf{r}\right)
\exp\left(  -iE_{i}t\right)  . \label{e}%
\end{equation}
It is convenient to replace the negative energy $E_{i}$ of the bound state by
its positive equivalent binding energy%
\begin{equation}
E_{B}=-E_{i}, \label{f}%
\end{equation}
which is often called the ionization potential $IP$ or $I_{p}$.

For laser applications, it is suitable to use the single-frequency form of the
Volkov solution. For linear polarization of the laser, this is%
\begin{equation}
\Psi^{Volk}=\frac{1}{V^{1/2}}\exp\left[  i\left(  \mathbf{p\cdot r}%
-\frac{p^{2}}{2}t-z\omega t+\zeta^{lin}\sin\omega t-\frac{z}{2}\sin2\omega
t\right)  \right]  , \label{l}%
\end{equation}
associated with a field given as%
\begin{equation}
\mathbf{A}=\mathbf{A}_{0}\cos\omega t, \label{m}%
\end{equation}
and where the parameters are introduced that%
\begin{equation}
\zeta^{lin}=A_{0}p\cos\theta/\omega c,\quad z=U_{p}/\omega,\quad U_{p}=\left(
A_{0}/2\omega\right)  ^{2}, \label{n}%
\end{equation}
where $\theta$ is the angle between the momentum vector $\mathbf{p}$ and the
polar axis of spherical coordinates, taken to be in the polarization direction
of the laser field. The quantity $U_{p}$ is the ponderomotive energy of the
electron in the laser field. A convenient way to proceed is to note that Eq.
(\ref{l}) includes the generating function for the Generalized Bessel Function
(GBF), which is%
\begin{equation}
\exp\left[  i\left(  \zeta^{lin}\sin\omega t-\frac{z}{2}\sin2\omega t\right)
\right]  =\sum_{n=-\infty}^{\infty}e^{in\omega t}J_{n}\left(  \zeta
^{lin},-\frac{z}{2}\right)  . \label{o}%
\end{equation}
The GBF can be represented either in an integral form or a summation form:%
\begin{equation}
J_{n}\left(  u,v\right)  =\frac{1}{2\pi}\int_{-\pi}^{\pi}d\varphi\exp\left[
i\left(  u\sin\varphi+v\sin2\varphi-n\varphi\right)  \right]  =\sum
_{k=-\infty}^{\infty}J_{n-2k}\left(  u\right)  J_{k}\left(  v\right)  ,
\label{p}%
\end{equation}
where $J_{n}\left(  x\right)  $ is an ordinary Bessel function of the first
kind. An extensive listing of properties of the GBF is in Appendices B-D of
Ref. \cite{hr80}.

The differential transition rate for atomic ionization by linearly polarized
laser light is, in atomic units%
\begin{align}
\frac{dW}{d\Omega} &  =\frac{1}{\left(  2\pi\right)  ^{2}}\sum_{n=n_{0}%
}^{\infty}p\left(  \frac{p^{2}}{2}+E_{B}\right)  ^{2}\left\vert \widehat{\phi
}_{i}\left(  \mathbf{p}\right)  \right\vert ^{2}\left[  J_{n}\left(
\zeta^{lin},-\frac{z}{2}\right)  \right]  ^{2},\label{q}\\
\frac{p^{2}}{2} &  =n\omega-U_{p}-E_{B},\quad n_{0}=\left\lceil U_{p}%
+E_{B}\right\rceil ,\label{r}\\
\widehat{\phi}_{i}\left(  \mathbf{p}\right)   &  =\int_{-\infty}^{\infty}%
d^{3}re^{-i\mathbf{p\cdot r}}\phi_{i}\left(  \mathbf{r}\right)  ,\label{s}%
\end{align}
where $\Omega$ is the solid angle in spherical coordinates with the
polarization direction of the electric field as the polar axis; the symbol
$\left\lceil ...\right\rceil $ in Eq. (\ref{r}) is the \textit{ceiling
function}, meaning the smallest integer containing the quantity within the
brackets. This assures that the energy conservation condition in Eq. (\ref{r})
always yields a final kinetic energy $p^{2}/2$ that is positive; and attention
is drawn to the definition of the momentum wave function in Eq. (\ref{s}),
that does not employ the $\left(  2\pi\right)  ^{-3/2}$ factor often found.

The result for photoionization by a circularly polarized laser field is%
\begin{align}
\frac{dW}{d\Omega}  &  =\frac{1}{\left(  2\pi\right)  ^{2}}\sum_{n=n_{0}%
}^{\infty}p\left(  \frac{p^{2}}{2}+E_{B}\right)  ^{2}\left\vert \widehat{\phi
}_{i}\left(  \mathbf{p}\right)  \right\vert ^{2}\left[  J_{n}\left(
\zeta^{circ}\right)  \right]  ^{2},\label{t}\\
\zeta^{circ}  &  =A_{0}p\sin\theta, \label{u}%
\end{align}
where the spherical coordinates are now defined with respect to the
propagation direction of the laser field, and Eqs. (\ref{r}) and (\ref{s})
remain unchanged.

It is instructive to note that an alternative way to write the GBF in Eq.
(\ref{q}) is%
\begin{equation}
J_{n}\left(  \zeta^{lin},-\frac{z}{2}\right)  =J_{n}\left(  \alpha_{0}%
^{lin}p\cos\theta,-\beta_{0}c\right)  , \label{v}%
\end{equation}
where $\alpha_{0}^{lin}$ is the amplitude in the electric-field direction of
the \textquotedblleft figure-8\textquotedblright\ motion of a charged particle
in a plane-wave field, as expressed in the frame of reference where the
particle is on average at rest, and $\beta_{0}$ is the amplitude of the
figure-8 in the propagation direction. The information thus conveyed is that
the magnetic field effect required to produce the figure-8 is inherent in the
Volkov solution. This is lost in the dipole approximation, where a magnetic
field cannot exist. This conclusion applies also to the case of circular
polarization, where the Bessel function can be written as%
\begin{equation}
J_{n}\left(  \zeta^{circ}\right)  =J_{n}\left(  \alpha_{0}^{circ}p\sin
\theta\right)  , \label{w}%
\end{equation}
where $\alpha_{0}^{circ}$ is the radius of the circular orbit followed by a
charged particle in a circularly polarized laser field as expressed in the
coordinate system where the center of the circular orbit is at rest. These
concepts depend on the notion of a propagation direction of the laser field,
whereas propagation is nonexistent in the dipole approximation.

Integration over solid angle makes use of symmetry about the polar axis to
give%
\begin{equation}
\int d\Omega...=2\pi\int d\left(  \cos\theta\right)  ...=2\pi\int_{-1}%
^{1}dx...=4\pi\int_{0}^{1}dx...,\label{w1}%
\end{equation}
and the complete ionization rate is%
\begin{equation}
W=\frac{1}{\pi}\sum_{n}\int_{0}^{1}dxp\left(  \frac{p^{2}}{2}+E_{B}\right)
^{2}\left\vert \widehat{\phi}_{i}\left(  \mathbf{p}\right)  \right\vert
^{2}J_{n}^{2},\label{w2}%
\end{equation}
where $J_{n}$ can be either the GBF of linear polarization as in Eq. (\ref{q})
or the ordinary Bessel function for circular polarization as in Eq. (\ref{t}).

In practical application of the above results, the squared momentum wave
function is of basic importance. This is easily found analytically for
hydrogenic states. Some sample results are:%
\begin{align}
\left\vert \widehat{\phi}_{i}\left(  \mathbf{p}\right)  \right\vert
_{H1s}^{2}  &  =\frac{4\pi}{\left(  p^{2}/2+E_{B}\right)  ^{4}}\label{x}\\
\left\vert \widehat{\phi}_{i}\left(  \mathbf{p}\right)  \right\vert
_{H2s}^{2}  &  =\frac{\pi}{2}\frac{\left(  p^{2}/2-E_{B}\right)  ^{2}}{\left(
p^{2}/2+E_{B}\right)  ^{6}}\label{y}\\
\left\vert \widehat{\phi}_{i}\left(  \mathbf{p}\right)  \right\vert
_{H2p}^{2}  &  =\frac{\pi}{24}\frac{p^{2}}{\left(  p^{2}/2+E_{B}\right)  ^{6}%
}. \label{z}%
\end{align}

The noble gases are often used in laser experiments. Analytical Hartree-Fock
wave functions are used according to the conventions of Radzig and Smirnov
\cite{radsmir}. It is convenient to define $PHIP$ as that part of the
momentum-space wave function that has all multiplicative factors extracted,
with only Hartree-Fock factors remaining. The result for helium is%
\begin{align}
W  &  =16\sum_{n}\int_{0}^{1}dxp\left(  \frac{p^{2}}{2}+E_{B}\right)
^{2}\left\vert PHIP\right\vert ^{2}J_{n}^{2},\label{aa}\\
PHIP  &  =\sum_{i=1}^{5}\frac{C_{i}N_{i}\zeta_{i}}{\left(  p^{2}+\zeta_{i}%
^{2}\right)  ^{2}}, \label{ab}%
\end{align}
where the $C_{i},N_{i},\zeta_{i}$ Hartree-Fock coefficients for helium are
given on page 62 of Ref. \cite{radsmir}.

Neon, argon, krypton, and xenon all have initial p-states. After averaging
over initial substates, all have the same general form. The average over
initial substates gives%
\begin{align}
\frac{1}{3}\left(  Y_{1}^{1}+Y_{0}^{1}+Y_{-1}^{1}\right)   &  =\frac{1}%
{\sqrt{24\pi}}\left(  \sqrt{2}\cos\theta-2i\sin\theta\sin\varphi\right)
,\label{ac}\\
\left\vert \widehat{\phi}_{i}\left(  \mathbf{p}\right)  \right\vert ^{2}  &
=\frac{256\pi}{3}\left(  \cos^{2}\theta+2\sin^{2}\theta\sin^{2}\varphi\right)
p^{2}\left\vert PHIP\right\vert ^{2},\label{ad}\\
\int d\Omega\left(  \cos^{2}\theta+2\sin^{2}\theta\sin^{2}\varphi\right)   &
=4\pi, \label{ae}%
\end{align}
where $\theta$ is the polar angle and $\varphi$ is the azimuthal angle of
spherical coordinates. This gives the general form of the ionization rate as%
\begin{equation}
W=\frac{256}{3}\sum_{n=n_{0}}^{\infty}\int_{0}^{1}dxp^{3}\left(  \frac{p^{2}%
}{2}+E_{B}\right)  ^{2}\left\vert PHIP\right\vert ^{2}J_{n}^{2}, \label{af}%
\end{equation}
when defining $x=\cos\theta$ as in Eq. (\ref{w1}). The necessary values of
$PHIP$ from Ref. \cite{radsmir} are, with appropriate Radzig and Smirnov page
references:%
\begin{align}
\text{Neon, p.63}  &  \text{: \ }PHIP=\sum_{i=1}^{4}\frac{C_{i}N_{i}\zeta_{i}%
}{\left(  p^{2}+\zeta_{i}^{2}\right)  ^{3}},\label{ag}\\
\text{Argon, p.64}  &  \text{: \ }PHIP=\sum_{i=1}^{2}\frac{C_{i}N_{i}\zeta
_{i}}{\left(  p^{2}+\zeta_{i}^{2}\right)  ^{3}}+\sum_{i=3}^{4}\frac{C_{i}%
N_{i}\left(  5\zeta_{i}^{2}-p^{2}\right)  }{\left(  p^{2}+\zeta_{i}%
^{2}\right)  ^{4}},\label{ah}\\
\text{Krypton, p.67}  &  \text{: \ }PHIP=\sum_{i=1}^{2}\frac{C_{i}N_{i}\left(
5\zeta_{i}^{2}-p^{2}\right)  }{\left(  p^{2}+\zeta_{i}^{2}\right)  ^{4}}%
+\sum_{i=3}^{4}\frac{6C_{i}N_{i}\left(  5\zeta_{i}^{2}-3p^{2}\right)
}{\left(  p^{2}+\zeta_{i}^{2}\right)  ^{5}},\label{ai}\\
\text{Xenon, p.70}  &  \text{: }PHIP=\frac{C_{1}N_{1}\left(  5\zeta_{1}%
^{2}-p^{2}\right)  }{\left(  p^{2}+\zeta_{1}^{2}\right)  ^{4}}+\sum_{i=2}%
^{3}\frac{6C_{i}N_{i}\zeta_{i}\left(  5\zeta_{i}^{2}-3p^{2}\right)  }{\left(
p^{2}+\zeta_{i}^{2}\right)  ^{5}}\label{aj}\\
&  +\text{\ }\sum_{i=4}^{5}\frac{6C_{i}N_{i}\left(  35\zeta_{i}^{4}%
-42\zeta_{i}^{2}p^{2}+3p^{4}\right)  }{\left(  p^{2}+\zeta_{i}^{2}\right)
^{6}}. \label{ak}%
\end{align}

The results for the transition rate are used to represent local conditions in
a numerical integration over the spatial and temporal distribution of the
field intensity in the laser focus. In the course of this integration, the
information necessary for angular distributions, momentum distributions, and
energy spectra can be extracted. If it is expected that saturation effects
might occur, then it is necessary to use the transition rate information in
the context of the solution of a rate equation. This procedure is described in
Section 4.3.6 of Ref. \cite{hrrev}.